\documentclass[useAMS,usenatbib]{mn2e}

\usepackage{graphicx}
\usepackage{subfigure}
\usepackage{rotating} 
\usepackage{float}

\newcommand{\lsim }{{\lower0.8ex\hbox{$\buildrel <\over\sim$}}}
\newcommand{\gsim }{{\lower0.8ex\hbox{$\buildrel >\over\sim$}}}

\usepackage{soul}

\def\Chandra{\emph{Chandra}}

\def\Swiftxrt{\emph{Swift/XRT}}

\def\simge{\mathrel{%
  \rlap{\raise 0.511ex \hbox{$>$}}{\lower 0.511ex \hbox{$\sim$}}}}
\def\simle{\mathrel{
  \rlap{\raise 0.511ex \hbox{$<$}}{\lower 0.511ex \hbox{$\sim$}}}}

\newcommand{\Msun}{\ifmmode {M_{\odot}}\else${M_{\odot}}$\fi}
\newcommand{\Lsun}{\ifmmode {L_{\odot}}\else${L_{\odot}}$\fi}
\newcommand{\Rsun}{\ifmmode {R_{\odot}}\else${R_{\odot}}$\fi}

\usepackage{xargs}
\usepackage{color}

\title[A new symbiotic X-ray binary]{Revealing a new symbiotic X-ray binary with Gemini NIFS}
\author[Bahramian et al.]
{Arash Bahramian$^{1}$\thanks{E-mail: bahramia@ualberta.ca}, Jeanette C. Gladstone$^{1}$, Craig O. Heinke$^{1}$, Rudy Wijnands$^{2}$,  \newauthor Ramanpreet Kaur$^{3}$, Diego Altamirano$^{4}$\\
$^{1}$ Dept. of Physics, University of Alberta, CCIS 4-183, Edmonton, Alberta, T6G 2E1, Canada\\
$^{2}$ Astronomical Institute 'Anton Pannekoek', University of Amsterdam, Science Park 904, 1098 XH, Amsterdam, the Netherlands\\
$^{3}$ Physics Department, Suffolk University, 41 Temple Street, Boston, Massachusetts, 02114, USA\\
$^{4}$ Physics \& Astronomy, University of Southampton, Southampton, Hampshire, SO17 1BJ, UK}
\begin{document}

\date{}

\pagerange{\pageref{firstpage}--\pageref{lastpage}} \pubyear{2014}

\maketitle

\label{firstpage}

\begin{abstract}
We use K-band spectroscopy of the counterpart to the rapidly variable X-ray transient XMMU J174445.5-295044 to identify it as a new symbiotic X-ray binary.  XMMU J174445.5-295044 has shown a hard X-ray spectrum (we verify its association with an Integral/IBIS 18-40 keV detection in 2013 using a short Swift/XRT observation), high and varying $N_H$, and rapid flares on timescales down to minutes, suggesting wind accretion onto a compact star. We observed its near-infrared counterpart using the Near-infrared Integral Field Spectrograph (NIFS) at Gemini-North, and classify the companion as $\sim$M2 III. We infer a distance of $3.1^{+1.8}_{-1.1}$ kpc (conservative 1$\sigma$ errors), and therefore calculate that the observed X-ray luminosity (2-10 keV) has reached to at least 4$\times10^{34}$ erg s$^{-1}$. We therefore conclude that the source is a symbiotic X-ray binary containing a neutron star (or, less likely, black hole) accreting from the wind of a giant.
\end{abstract}


\section{Introduction}
Symbiotic binaries transfer mass via the winds of cold (usually late K or M) giants onto compact objects: white dwarfs, neutron stars or black holes \citep{Kenyon86}, with orbital periods typically in the 100s to 1000s of days \citep{Belczynski00}.  They were first identified by the presence of high-ionization emission lines in optical spectra of otherwise cold giants, indicating the presence of two components of vastly different temperatures.  ROSAT X-ray studies of symbiotic binaries distinguished three classes ($\alpha$, $\beta$, $\gamma$) by the X-ray spectral shape \citep{Murset97}, with higher energy X-ray measurements adding two further classes showing highly-absorbed spectra \citep{Luna13}.  A small but rapidly increasing number of symbiotic systems have been identified as containing a neutron star as an accretor, through the measurement of pulsations and/or hard X-ray emission above 20 keV, and are known as symbiotic X-ray binaries \citep{Masetti06}.  

Only seven symbiotic X-ray binaries have been positively identified so far; GX 1+4, \citep{Davidsen77}; 4U 1700+24, \citep{Masetti02}; 4U 1954+319, \citep{Masetti06}; Sct X-1, \citep{Kaplan07}; IGR J16194-2810, \citep{Masetti07}; IGR J16358-4726, \citep{Nespoli10}; and XTE J1743-363, \citep{Bozzo13}.  Several other likely candidate systems have also been proposed (e.g. \citealt{Nucita07}, \citealt{Masetti11}, \citealt{Hynes14}).  The identification and characterization of a symbiotic X-ray binary requires clear information on the nature of the accretor (e.g. from pulsations or unusual luminosities) and the donor (e.g. from spectroscopy).  

\citet{Heinke09c} identified XMMU J174445.5-295044 as a rapidly variable (timescales down to 100s of seconds) Galactic transient, using nine \emph{XMM-Newton}, \emph{Chandra}, and \emph{Suzaku} observations. It showed 2-10 keV X-ray fluxes up to $>3\times10^{-11}$ erg cm$^{-2}$ s$^{-1}$, and variations in $N_H$, from $8\times10^{22}$ up to $15\times10^{22}$ cm$^{-2}$.  The rapid variations and variable $N_H$ suggested accretion from a clumpy wind, rather than an accretion disk. \citet{Heinke09c} also identified a bright near-infrared (NIR) counterpart (2MASS J17444541-2950446) within the 2'' XMM error circle.  \citet{Heinke09c} calculated the probability of a star of this brightness in $K_S$ appearing in the X-ray error circle as only 2\%, indicating that it is almost certainly the true counterpart.  This star appears highly obscured and shows infrared colors typical of late-type stars, which Heinke et al. suggested indicates that XMMU J174445.5-295044 is a symbiotic star or symbiotic X-ray binary.  

The \emph{INTEGRAL} Galactic bulge monitoring program \citep{Kuulkers07} reported an X-ray transient detected by the JEM-X monitor on March 23, 2012 \citep{Chenevez12}, at 17:44:48, -29:51:00, with an uncertainty of 1.3$'$ at 95\% confidence, consistent with XMMU J174445.5-295044.  The 10-25 keV flux of $1.5 \pm 0.3 \times 10^{-10}$ erg cm$^{-2}$ s$^{-1}$ is larger than previously reported for XMMU J174445.5-295044, but the high estimated $N_H$ (not specified, but the JEM-X source was undetected below 10 keV, indicating $N_H$$>$$10^{23}$ cm$^{-2}$) suggests that this is likely the same source, as it is known to exhibit similarly large intrinsic extinction \citep{Heinke09c}.  In March 2013, the INTEGRAL IBIS telescope detected a hard transient at $9.3\pm1.4\times10^{-11}$ erg cm$^{-2}$ s$^{-1}$ (17-60 keV), at position 17:44:41.76, -29:48:18.0, uncertainty 4.2' \citep{Krivonos13}.  Krivonos et al. note that this position is consistent with XMMU J174445.5-295044, but suggest that follow-up observations are needed to verify whether it is the same source. 

In this paper, we present Gemini NIFS spectroscopy of 2MASS J17444541-2950446, and conclude that its spectral type indicates a M2 III giant. We also describe a Swift/XRT observation permitting the confident identification of the 2013 INTEGRAL/IBIS transient \citep{Krivonos13} with XMMU J174445.5-295044.  We combine these results to infer a peak $L_X$ $>$ 4 $\times10^{34}$ erg s$^{-1}$. These results together allow us to confidently identify XMMU J174445.5-295044 as a symbiotic X-ray binary containing a neutron star or black hole accretor, rather than a white dwarf.

\section{Data and Analysis}\label{sec_data}
\subsection{Swift/XRT}\label{sec_xray}

We observed XMMU J174445.5-295044 with \emph{Swift/XRT} on March 30th, 2013 (two days after the INTEGRAL/IBIS detection of \citealt{Krivonos13}) in photon-counting mode. The observation was interrupted after an on-source exposure of $\sim150$ seconds due to a Gamma-ray burst alert.
We detect a single source in the 23.6' diameter field, showing 7 counts.
Using FTOOLS \emph{xrtcentroid} we determine the position to be RA = 17:44:46.26s and DEC = -29:50:56.08$''$ with positional uncertainty of 9$''$,  consistent within $<$2$\sigma$ with the position of XMMU J174445.5-295044.  Given the lack of other X-ray sources nearby (see figure 4 of \citealt{Heinke09c}), the Swift/XRT source is thus certainly the same as XMMU J174445.5-295044.

\begin{figure*}
 \includegraphics[scale=0.6]{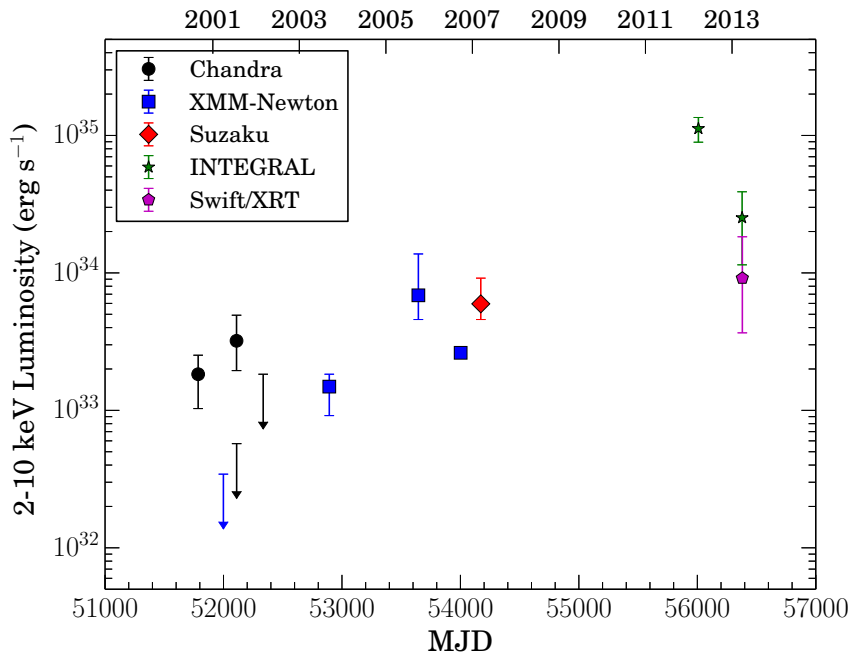}
 \caption{ Long term X-ray light curve of XMMU J174445.5-295044, calculating X-ray luminosities in the 2-10 keV band using our calculated 3.1 kpc distance (\S~\ref{distance}). Errorbars do not include the distance uncertainty, so that intrinsic variations can be more clearly seen. \Chandra\ (black circles \& upper limits), \emph{XMM-Newton} (blue squares \& upper limits) and \emph{Suzaku} (red diamonds) fluxes  from \citet{Heinke09c}. \emph{INTEGRAL} 2-10 keV fluxes (green stars) are extrapolated from the fluxes reported by \citet[JEM-X, 10-25 keV]{Chenevez12} and  \citet [IBIS/ISGRI, 17-60 keV]{Krivonos13}. Our \Swiftxrt\ observation (magenta pentagon) is reported in \S~\ref{sec_xray}. The final \emph{XMM-Newton datapoint has statistical errorbars smaller than the marker size.}}
 \label{fig_xray_lc}
\end{figure*}

We reprocessed the Swift/XRT data (using HEASOFT 6.14), extracted a spectrum with XSELECT, and created an effective area file with \emph{xrtmkarf}.  
We fit the Swift/XRT spectrum with an absorbed power-law using \emph{cstat} statistics \citep{Cash79} in XSPEC 12.8.1, fixing the photon index to 1.18 (as found in the deepest and most-constrained observation of \citealt{Heinke09c}). We measure $N_H=4.5_{-3.3}^{+7.2}\times10^{22}$ cm$^{-2}$ and an unabsorbed 2-10 keV flux of 8.5$_{-5.0}^{+10.3}\times10^{-12}$ erg s$^{-1}$ cm$^{-2}$, the second-highest 2-10 keV flux recorded from XMMU J174445.5-295044.

If the INTEGRAL/IBIS measurement is extrapolated (using the spectrum above) to the 2-10 keV band, the Swift/XRT flux measurement is 1/3 of the INTEGRAL measurement.  Such a high flux from XMMU J174445.5-295044 only two days after the INTEGRAL/IBIS detection, combined with the lack of other detections within the 4.2$'$ INTEGRAL/IBIS error circle, indicates that XMMU J174445.5-295044 was the origin of the INTEGRAL/IBIS detection.  The 2012 INTEGRAL/JEM-X detection \citep{Chenevez12}, with a smaller error circle of 1.3$'$, can also be confidently assigned to XMMU J174445.5-295044.

We created long term X-ray lightcurves of XMMU J174445.5-295044, calculating X-ray luminosities in the 2-10 keV band using our calculated 3.1 kpc distance (see below). 
In Figure~\ref{fig_xray_lc}, we show the published detections of XMMU J174445.5-295044 (including our Swift measurement, and associating the two INTEGRAL detections, extrapolating their flux down to 2-10 keV). 

\subsection{Infrared data}
\subsubsection{Data and reduction}
We observed XMMU J174445.5-295044 with Near-infrared Integral Field Spectrograph (NIFS, \citealt{McGregor02}) mounted on the Fredrick C. Gillett telescope at Gemini-North observatory. The observation was done in queue mode on July 9th, 2012 under program ID GN-2012A-Q-114 (PI: C.~O.~Heinke). NIFS provides spectroscopy with spectral resolving power R$\sim$ 5000 over a 3.0$''\times$3.0$''$ field of view in the Z through K-band (9500 to 24000 \AA). We performed the observation in K-band with standard methods for near-infrared, with a series of observations pointing on-source and blank sky. Blank sky observations were done in order to subtract sky emission from on-source observations. In order to remove telluric features in the spectrum of our target, we observed the A0V star HIP 88566 at similar airmass. For wavelength calibration, an exposure of Argon/Xenon arc lamps was taken. Also for spatial distortion removal and calibration, exposures with a Ronchi mask were taken.

We reduced and reprocessed the data using Gemini IRAF package V1.12 beta 2 included in IRAF\footnote{IRAF is distributed by the National Optical Astronomy Observatory, which is operated by the Association of Universities for Research in Astronomy (AURA) under cooperative agreement with the National Science Foundation.} (V2.16) distributed in Ureka\footnote{Ureka is provided by Association of Universities for Research in Astronomy (AURA).} 1.0 beta 5. NIFS package contains recipes for three stages of data reduction (baseline calibration, telluric data and science data) in ``\emph{nifsexamples}''\footnote{http://www.gemini.edu/sciops/instruments/nifs/?q=node/10356}. In baseline calibration we made flat field and bad pixel map, performed wavelength calibration and determined the spatial curvature and spectral distortion in the Ronchi flat.

The spectra of A0V stars in the K band only show one significant feature, at Br$\gamma$ (21661 \AA). We removed this stellar feature from our reference spectrum to obtain a pure telluric spectrum. After extracting the one-dimensional spectrum of the telluric star, we divided the spectrum with a blackbody spectrum and included a Voigt profile fit to the Br$\gamma$ feature (following \citealt{Barbosa08}) to mimic the A0V spectrum of our calibration source. This spectrum was created using \emph{mk1dspec} in \emph{artdata} package. We assumed a temperature of $\sim$9800 K for the blackbody continuum of the A0V telluric star \citep{Adelman04} and determined the Voigt profile parameters by fitting with the task \emph{splot}. We then eliminated telluric features in the science spectrum using the achieved pure telluric spectrum with the task \emph{nftelluric}. The final output of the reduction stage is calibrated telluric-corrected data in the form of a three-dimensional data cube with two spatial dimensions, each 62 pixels wide, and one spectral (wavelength) dimension of 2040 pixels. We extracted a one-dimensional spectrum by merging spatial dimensions inside a circular region with radius of 7 pixels centred on the source using DS9. 

\subsubsection{Spectral analysis}\label{sec:spec}
\begin{figure*}
 \includegraphics[scale=0.42]{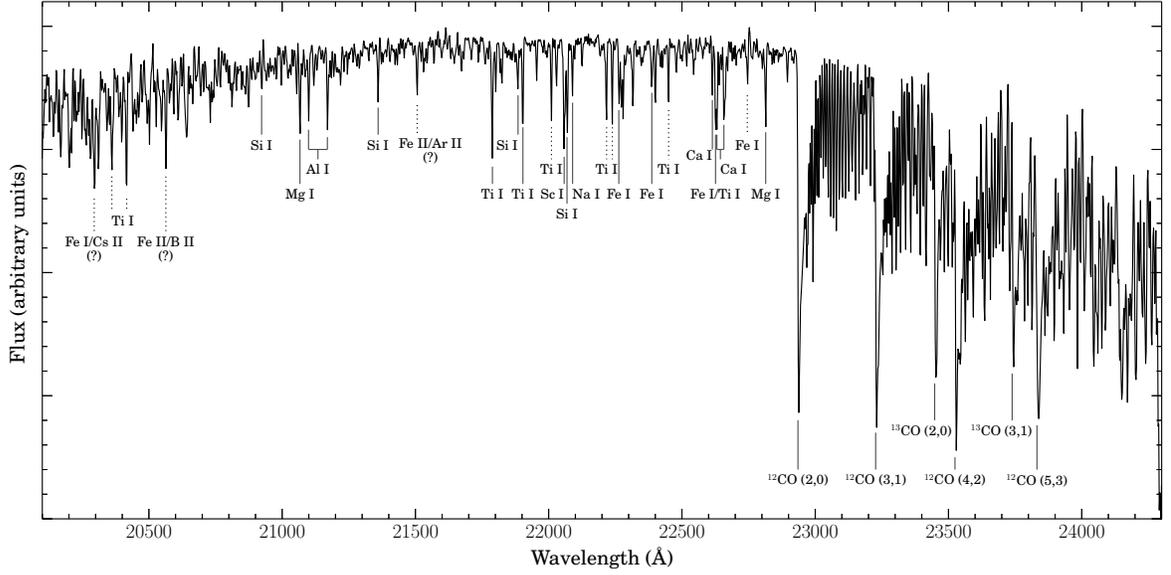}
 \caption{K-band spectrum of XMMU 174445.5-295044. Identified line profiles are listed in Table~\ref{tab_lines}. Features identified using spectral libraries \citep{Kleinmann86, Ramirez97} are marked with solid lines while features identified using NIST/ASD are marked with dashed lines. Ambiguous and uncertain identifications are labeled with ``?".}
 \label{fig_spectrum}
\end{figure*}

We measured the radial velocity of the source, using the \emph{rvidlines} task in IRAF RV package to achieve a red/blue shift-corrected spectrum. In order to do this we first needed to identify a small number of prominent lines in the spectrum of this source. These include Al I (21170 \AA), Si I (21360 \AA), Ti (21789 \AA, 21903 \AA), Na I (22090 \AA) and Ca I (22614 \AA). \emph{rvidlines} provided us with a velocity correction of -12$\pm$3 km s$^{-1}$, which was applied to the full spectrum. This corrected spectrum can be seen in Figure~\ref{fig_spectrum}. 

We used various available spectral libraries for late-type stars \citep{Kleinmann86, Wallace97, Ramirez97} to identify spectral features present in the spectrum. To obtain accurate identification and vacuum wavelength values we compared these identifications with data available in National Institute of Standards and Technology Atomic Spectra Database (NIST-ASD, \citealt{Kramida13}). Figure~\ref{fig_spectrum} shows the rest-frame spectrum of XMMU J174445.5-295044 with all identified features. These features are tabulated in Table~\ref{tab_lines}. 

\begin{table}
 \centering
  \caption{Identified spectral lines in the spectrum. Reported wavelengths are in rest frame. References: 1- \citet{Kleinmann86}, 2- \citet{Ramirez97}, 3- NIST-ASD. ``?" indicates uncertain identifications.}
  \begin{tabular}{@{}lcc@{}}
  \hline
  Species	&	Wavelength(\AA)	& Reference\\
  \hline
  Fe I/Cs II	(?)&	20295	&	3\\
  Ti I		&	20361	&	3\\
  Fe II/B II (?) &	20563	&	3\\
  Si I		&	20923	&	1,3\\
  Mg I	&	21067	&	1,3\\
  Al I		&	21170	&	1,3\\
  Si I		&	21360	&	1\\
  Fe II/Ar II (?)	&	21506	&	3\\
  Ti I		&	21789	&	1,3\\
  Si I		&	21885	&	1\\
  Ti I		&	21903	&	1,3\\
  Ti I		&	22010	&	3\\
  Sc I		&	22058	&	2,3\\
  Si I		&	22069	&	2,3\\
  Na I	&	22090	&	1,3\\
  Ti I		&	22217	&	3\\
  Fe I		&	22263	&	1,3\\
  Fe I		&	22387	&	1,3\\
  Ti I		&	22450	&	3\\
  Ca I	&	22614	&	1,3\\
  Fe I		&	22626	&	2,3\\
  Ti I		&	22627	&	2,3\\
  Ca I	&	22657	&	1,3\\
  Fe I		&	22745	&	3\\
  Mg I	&	22814	&	1,3\\
  $^{12}$CO (2,0)	&	22935	&	1\\
  $^{12}$CO (3,1)	&	23227	&	1\\
  $^{13}$CO (2,0)	&	23448	&	1\\
  $^{12}$CO (4,2)	&	23524	&	1\\
  $^{13}$CO (3,1)	&	23739	&	1\\
  $^{12}$CO (5,3)	&	23832	&	1\\
  \hline
\end{tabular}
\label{tab_lines}
\end{table}

\begin{figure*}
\begin{center}
 \includegraphics[scale=0.42]{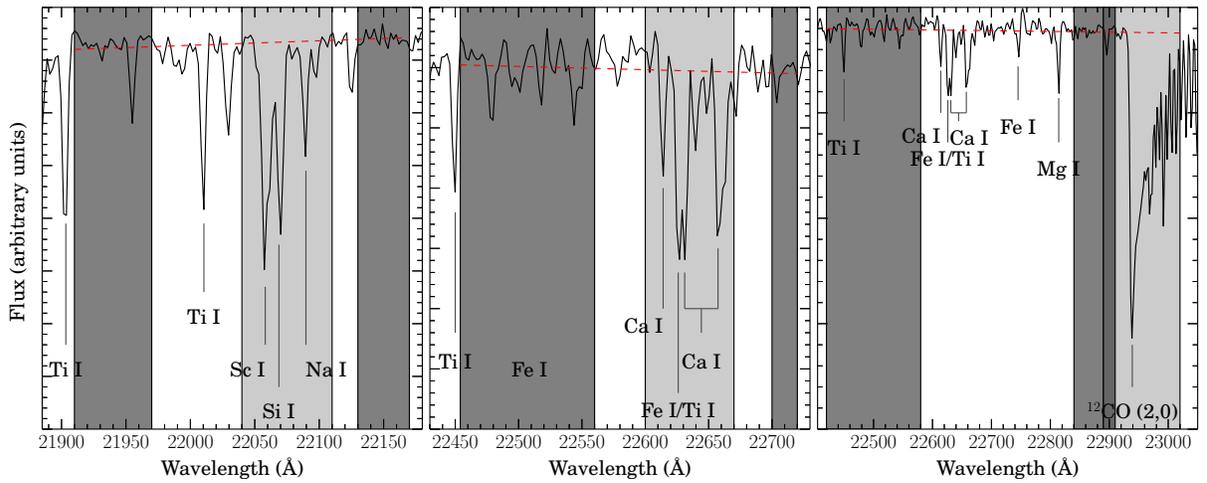}
 \caption{Chosen features and continua intervals used to obtain the spectral classification of the companion in this system. Left to right: Na I, Ca I, $^{12}$CO(2,0). Light shaded regions show chosen regions for features and dark shaded regions represent chosen continua regions. These regions are tabulated in Table~\ref{tab_features}. The dashed lines represent interpolated continuum level in each case.}
 \label{fig_eqw}
\end{center}
\end{figure*}

For two-dimensional stellar classification (spectral and luminosity) of the source we followed the method discussed by \citet{Ramirez97, Ivanov04, Comeron04}. This method consists of comparing the strength of a feature which is temperature-dependent with a feature which is temperature- and surface gravity-dependent. 
Following the method outlined in \citet{Comeron04},  we selected wavelength regions encompassing significant temperature-dependent features (Na I, Ca I), and one region representing a temperature- and surface gravity-dependent feature ($^{12}$CO). 
For each feature, we use two nearby, featureless continuum regions to approximate the expected continuum level within the feature by linear interpolation. We used nearly the same feature and continuum definitions as \citet{Comeron04} (Table~\ref{tab_features}). We made a small modification to the range of the blue continuum region for the Ca I feature given by \citet{Comeron04}, shortening it by 4 \AA\ to avoid including the relatively strong nearby Ti I (~22450 \AA) line. This modification has an effect of $<0.2$ $\%$ on the equivalent width measurement. These features and continua are represented in Figure~\ref{fig_eqw}. Finally we calculated an equivalent width for each feature, and compared these values to the values reported in \citet{Comeron04}. 

To estimate errors, we divided the continuum regions into halves, and computed the equivalent widths using either half alone.  We took the largest variation from our reported values as an estimate of the error in each measurement.

\section{Results and Discussion}\label{sec_discus}

\subsection{Two-dimensional spectral classification}
\citet{Comeron04} demonstrate that the $^{12}$CO feature for supergiants always shows an equivalent width (EW) of $>25$~{\AA} (see their figures 8-13). We obtained EW[$^{12}$CO]$\approx19.4\pm0.1$~\AA, which is typical for giants or dwarfs \citep{Comeron04}. Thus we can rule-out the possibility of a supergiant.

 \citet{Ramirez97} and \citet{Ivanov04} show that $\log$[EW(CO)/(EW(Ca I)+EW(Na I))] can be used to separate giants from dwarfs.  \citet{Ramirez97} show that this quantity should be between -0.22 and 0.06 for dwarfs, vs.\ between 0.37 and 0.61 for giants. We found this quantity to be $0.67\pm{0.06}$ for our source, in agreement with the estimated range for giants. Presence of fairly strong $^{13}$CO bands in our spectrum is another indicator for a giant, as these features are invisible in a dwarf. 

To estimate the temperature of this source, we used the first-order relationship between effective temperature (T$_{eff}$) and EW[$^{12}$CO] (in angstroms) for giants proposed by \citet{Ramirez97}: 
\begin{equation}
T_{eff} = (5019\pm79)-(68\pm4)\times $EW$[^{12}$CO$]
\end{equation}
Considering the uncertainty in EW[$^{12}$CO], we found T$_{eff}$~=~3700~$\pm$~160 K. According to \citet{vanBelle99}, T$_{eff}$~=~3700 K indicates M2 giant; using \citet{Richichi99} suggests M1.5 while the relation in \citet{Ramirez97} gives an M1.7 giant. Thus, adopting either the van Belle or Richichi calibration, the resulting spectral type is M2 III, with a reasonable range from M0 to M3. If we used the less detailed calibration from Ramirez et al 1997, we obtain a similar result of M1.7 (M0 to M3).  Thus, we adopt M2 III as our spectral type, with a possible range from M0 to M3 III.

There is no evidence for a feature at Br$\gamma$ in our spectrum, either before or  after our telluric subtraction.  \citet{Nespoli10} see Br$\gamma$ emission from two symbiotic X-ray binaries. However, the similar P-Cygni shape of the Br$\gamma$ feature in both stars, and also in the supergiant X-ray binary IGR J16493-4348 studied by them with the same method, lend support to their hypothesis that this feature is a residual artifact of their telluric removal procedure (which is more complex than ours, involving using a G star as a second telluric reference).

\begin{table*}
 \centering
  \caption{Definition of spectral features, chosen continuum intervals and measured equivalent width. We used definitions in \citet{Comeron04} with a small modification to Ca I blue continuum (\S~\ref{sec:spec}).}
  \begin{tabular}{@{}cllllllc@{}}
  \hline
  		&		\multicolumn{2}{c}{Band}			&		\multicolumn{2}{c}{Blue continuum}	&	\multicolumn{2}{c}{Red continuum}	& \\
  Feature	&	Center(\AA)	&	$\delta \lambda$	&	Center(\AA)	&	$\delta \lambda$	&	Center(\AA)	&	$\delta \lambda$	&	Equivalent width (\AA)\\
  \hline
  Na I	&	22075		&	70				&	21940		&	60				&	22150		&	40				&		2.23$\pm$0.14		\\
  Ca I	&	22635		&	110				&	22507		&	53				&	22710		&	20				&		1.93$\pm$0.40		\\
  $^{12}$CO &	22955		&	130				&	22500		&	160				&	22875		&	70				&		19.36$\pm$0.13	\\
  \hline
\end{tabular}
\label{tab_features}
\end{table*}

\subsection{Extinction, distance, and nature of the accretor}\label{distance}
We use our identification of the spectral type, with the 2MASS photometry \citep{Skrutskie06} reported by \citet{Heinke09c}, to estimate the extinction, and thus the distance, to XMMU J174445.5-295044, in a similar way as \citet{Kaplan07}, but explicitly accounting for the difference between the $K_S$ and $K$ bands.  Although the 2MASS colors were measured at a different time from the NIFS spectroscopy reported here, we do not expect large variations in the temperature or observed extinction of the giant, as the stars most affected by this are of later ($>$M5) spectral types \citep{Habing96}.

M2 III stars have an absolute magnitude of $M_J= -3.92$ and intrinsic $J$-$K_S$ colors of 1.12 \citep{Covey07}. \citet{Heinke09c} report a 2MASS magnitude of $m_J=14.89$ in $J$ for our object, and an observed $J$-$K_S$=4.72. 

We use $A_J/A_V$~=~0.282 \citep{Cardelli89}, and $A_J/A_{K_s}$~=~2.5$\pm0.2$ \citep{Indebetouw05}. Thus we infer $A_V$~=~$\frac{(J-K_S)_{obs}-(J-K_S)}{(A_J/A_V)-(A_Ks/A_V)}$~=~21.3$_{-0.1}^{+1.9}$, and $A_J$~=~6.0$_{-0.3}^{+0.5}$. The extinction measurement converts (using $N_H$ (cm$^{-2}$)~=~(2.21$\pm$ 0.09) $\times10^{21}$ $A_V$, \citealt{Guver09}), to $N_H~=~(4.7\pm0.5)\times10^{22}$ cm$^{-2}$, which is below the X-ray measured values (measurements of $8.6\pm0.4\times10^{22}$ cm$^{-2}$, and $16^{+5}_{-4}\times10^{22}$ cm$^{-2}$, from different observations) in \citet{Heinke09c}. This is consistent with expectations for a wind-accreting system, where much of the $N_H$ is expected to be local to the compact object, and with the evidence for variation in $N_H$ between different observations shown by \citet{Heinke09c}.

Using this $A_J$ estimate, the expected $M_J$ for a M2 III star, and the observed $J$ magnitude, we can thus estimate $d$=3.1 kpc as the most likely distance to our object.  The largest uncertainty in our distance estimate is our estimate of the absolute magnitude of the companion star.  Allowing for a conservative 1-magnitude uncertainty on the absolute magnitude (estimated from \citealt{Breddels10}; this is probably more precise than 1$\sigma$), we find $d$= $3.1^{+1.8}_{-1.1}$ kpc.  
This distance is consistent with our (small) radial velocity estimate, which would be typical of a disk star observed at a very small Galactic latitude ({\it l}~=~359.1$^\circ$), and with our measurement of the relative strengths of the CO and Na lines, the ratio of which is more consistent with disk giants than with giants in the bulge \citep{Comeron04}.   

From this distance estimate, we can infer the X-ray luminosities of XMMU J174445.5-295044, as plotted in Figure~\ref{fig_xray_lc} (errors there do not include the distance uncertainties).  The majority of the X-ray detections are between $10^{33}$ and $10^{34}$ erg s$^{-1}$, but the INTEGRAL/JEM-X detection in March 2012 \citep{Chenevez12} gives a (2-10 keV) X-ray luminosity of (1.1$\pm0.2)\times10^{35}$ erg s$^{-1}$ for $d$~=~3.1 kpc; even at the lower limit on the distance (d~=~2.0 kpc), the luminosity exceeds $4~\times~10^{34}$ erg s$^{-1}$ (Similarly, the March 2013 INTEGRAL/IBIS detection gives a (2-10 keV) $L_X~=~$2.5$~\times~10^{34}$ erg s$^{-1}$ for 3.1 kpc, or $1.1~\times~10^{34}$ erg s$^{-1}$ for the 2.0 kpc lower distance limit, which further confirms the high X-ray luminosity of XMMU J174445.5-295044).
Combining this high peak X-ray luminosity (four times the maximum seen for any accreting white dwarf, \citealt{Stacey11}) with the hard X-ray spectrum inferred from the later Integral/IBIS detection above 17 keV \citep{Krivonos13}, we can confidently rule out a white dwarf nature for the accretor.  Thus, we securely identify XMMU J174445.5-295044 as a symbiotic X-ray binary, with a neutron star (or, less likely, black hole) accreting from the wind of an M2 giant star.  

 XMMU J174445.5-295044 stands out from other symbiotic X-ray binaries only in not showing detectable X-ray pulsations \citep{Heinke09c}.   The complete absence of NIR spectroscopic evidence of accretion in our NIFS spectrum is typical of other symbiotic X-ray binaries with relatively low accretion rates.  The lack of detected pulsations also means that the accretor could be a black hole, though black hole symbiotic X-ray binaries should be less common.
 
 The increasing number of symbiotic binaries without detected emission lines in high-quality spectra being detected recently \citep{vandenBerg06,vandenBerg12,Hynes14}
 strongly suggests that there should be many more symbiotic stars (with white dwarf accretors) which also do not show optical/NIR spectroscopic evidence of accretion \citep{vandenBerg06}. Symbiotic systems may make up an important portion of the faint Galactic X-ray source population.

\section*{Acknowledgments}
Authors thank the anonymous referee for their helpful comments. C.O.H. thanks Jaehyon Rhee, Tim Davidge, and Inger Jorgensen for assistance in preparing the observing plan.  
A.B. thanks E. Rosolowsky for helpful discussions. We thank the Gemini helpdesk for helpful discussions on the analysis of NIFS data.  We acknowledge financial support from NSERC Discovery Grants (C.O.H.), an Alberta Ingenuity New Faculty Award (C.O.H.) and the Avadh Bhatia Fellowship (J.C.G.). RW and RK are supported by an European Research Council Starting Grant. DA acknowledges support from the Royal Society. \\

NIST-Atomic Spectra Database funded [in part] by the Office of Fusion Energy Sciences of the U.S. Department of Energy, by the National Aeronautics and Space Administration, by NIST's Standard Reference Data Program (SRDP), and by NIST's Systems Integration for Manufacturing Applications (SIMA) Program. The \Swiftxrt\ Data Analysis Software (XRTDAS) developed under the responsibility of the ASI Science Data Center (ASDC), Italy. We acknowledge extensive use of the ADS and arXiv.

\bibliographystyle{mn2e}
\bibliography{ref_list}

\begin{thebibliography}{}

\bibitem[\protect\citeauthoryear{{Adelman}}{{Adelman}}{2004}]{Adelman04}
{Adelman} S.~J.,  2004, in {Zverko} J.,  {Ziznovsky} J.,  {Adelman} S.~J.,
  {Weiss} W.~W.,  eds, The A-Star Puzzle Vol.~224 of IAU Symposium, {The
  physical properties of normal A stars}.
pp 1--11

\bibitem[\protect\citeauthoryear{{Barbosa}, {Blum}, {Conti}, {Damineli} \&
  {Figuer{\^e}do}}{{Barbosa} et~al.}{2008}]{Barbosa08}
{Barbosa} C.~L.,  {Blum} R.~D.,  {Conti} P.~S.,  {Damineli} A.,
  {Figuer{\^e}do} E.,  2008, \apjl, 678, L55

\bibitem[\protect\citeauthoryear{{Belczy{\'n}ski}, {Miko{\l}ajewska}, {Munari},
  {Ivison} \& {Friedjung}}{{Belczy{\'n}ski} et~al.}{2000}]{Belczynski00}
{Belczy{\'n}ski} K.,  {Miko{\l}ajewska} J.,  {Munari} U.,  {Ivison} R.~J.,
  {Friedjung} M.,  2000, \aaps, 146, 407

\bibitem[\protect\citeauthoryear{{Bozzo}, {Romano}, {Ferrigno}, {Campana},
  {Falanga}, {Israel}, {Walter} \& {Stella}}{{Bozzo} et~al.}{2013}]{Bozzo13}
{Bozzo} E.,  {Romano} P.,  {Ferrigno} C.,  {Campana} S.,  {Falanga} M.,
  {Israel} G.,  {Walter} R.,    {Stella} L.,  2013, \aap, 556, A30

\bibitem[\protect\citeauthoryear{{Breddels}, {Smith}, {Helmi}, {Bienaym{\'e}},
  {Binney}, {Bland-Hawthorn} \& {et al.}}{{Breddels} et~al.}{2010}]{Breddels10}
{Breddels} M.~A.,  {Smith} M.~C.,  {Helmi} A.,  {Bienaym{\'e}} O.,  {Binney}
  J.,  {Bland-Hawthorn} J.,    {et al.} 2010, \aap, 511, A90

\bibitem[\protect\citeauthoryear{{Cardelli}, {Clayton} \& {Mathis}}{{Cardelli}
  et~al.}{1989}]{Cardelli89}
{Cardelli} J.~A.,  {Clayton} G.~C.,    {Mathis} J.~S.,  1989, \apj, 345, 245

\bibitem[\protect\citeauthoryear{{Cash}}{{Cash}}{1979}]{Cash79}
{Cash} W.,  1979, \apj, 228, 939

\bibitem[\protect\citeauthoryear{{Chenevez}, {Kuulkers}, {Alfonso-Garzon},
  {Beckmann}, {Bird}, {Brandt}, {Del Santo} \& et al.}{{Chenevez}
  et~al.}{2012}]{Chenevez12}
{Chenevez} J.,  {Kuulkers} E.,  {Alfonso-Garzon} J.,  {Beckmann} V.,  {Bird}
  T.,  {Brandt} S.,  {Del Santo} M.,    et al. 2012, The Astronomer's Telegram,
  4000, 1

\bibitem[\protect\citeauthoryear{{Comer{\'o}n}, {Torra}, {Chiappini},
  {Figueras}, {Ivanov} \& {Ribas}}{{Comer{\'o}n} et~al.}{2004}]{Comeron04}
{Comer{\'o}n} F.,  {Torra} J.,  {Chiappini} C.,  {Figueras} F.,  {Ivanov}
  V.~D.,    {Ribas} S.~J.,  2004, \aap, 425, 489

\bibitem[\protect\citeauthoryear{{Covey}, {Ivezi{\'c}}, {Schlegel},
  {Finkbeiner}, {Padmanabhan}, {Lupton}, {Ag{\"u}eros}, {Bochanski}, {Hawley},
  {West}, {Seth}, {Kimball}, {Gogarten}, {Claire}, {Haggard}, {Kaib},
  {Schneider} \& {Sesar}}{{Covey} et~al.}{2007}]{Covey07}
{Covey} K.~R.,  {Ivezi{\'c}} {\v Z}.,  {Schlegel} D.,  {Finkbeiner} D.,
  {Padmanabhan} N.,  {Lupton} R.~H.,  {Ag{\"u}eros} M.~A.,  {Bochanski} J.~J.,
  {Hawley} S.~L.,  {West} A.~A.,  {Seth} A.,  {Kimball} A.,  {Gogarten} S.~M.,
  {Claire} M.,  {Haggard} D.,  {Kaib} N.,  {Schneider} D.~P.,    {Sesar} B.,
  2007, \aj, 134, 2398

\bibitem[\protect\citeauthoryear{{Davidsen}, {Malina} \& {Bowyer}}{{Davidsen}
  et~al.}{1977}]{Davidsen77}
{Davidsen} A.,  {Malina} R.,    {Bowyer} S.,  1977, \apj, 211, 866

\bibitem[\protect\citeauthoryear{{G{\"u}ver} \& {{\"O}zel}}{{G{\"u}ver} \&
  {{\"O}zel}}{2009}]{Guver09}
{G{\"u}ver} T.,  {{\"O}zel} F.,  2009, \mn, 400, 2050

\bibitem[\protect\citeauthoryear{{Habing}}{{Habing}}{1996}]{Habing96}
{Habing} H.~J.,  1996, \aapr, 7, 97

\bibitem[\protect\citeauthoryear{{Heinke}, {Tomsick}, {Yusef-Zadeh} \&
  {Grindlay}}{{Heinke} et~al.}{2009}]{Heinke09c}
{Heinke} C.~O.,  {Tomsick} J.~A.,  {Yusef-Zadeh} F.,    {Grindlay} J.~E.,
  2009, \apj, 701, 1627

\bibitem[\protect\citeauthoryear{{Hynes}, {Torres}, {Heinke}, {Maccarone},
  {Mikles}, {Britt}, {Knigge}, {Greiss}, {Jonker}, {Steeghs}, {Nelemans},
  {Bandyopadhyay} \& {Johnson}}{{Hynes} et~al.}{2014}]{Hynes14}
{Hynes} R.~I.,  {Torres} M.~A.~P.,  {Heinke} C.~O.,  {Maccarone} T.~J.,
  {Mikles} V.~J.,  {Britt} C.~T.,  {Knigge} C.,  {Greiss} S.,  {Jonker} P.~G.,
  {Steeghs} D.,  {Nelemans} G.,  {Bandyopadhyay} R.~M.,    {Johnson} C.~B.,
  2014, \apj, 780, 11

\bibitem[\protect\citeauthoryear{{Indebetouw}, {Mathis}, {Babler}, {Meade},
  {Watson}, {Whitney} \& et al.}{{Indebetouw} et~al.}{2005}]{Indebetouw05}
{Indebetouw} R.,  {Mathis} J.~S.,  {Babler} B.~L.,  {Meade} M.~R.,  {Watson}
  C.,  {Whitney} B.~A.,    et al. 2005, \apj, 619, 931

\bibitem[\protect\citeauthoryear{{Ivanov}, {Rieke}, {Engelbracht},
  {Alonso-Herrero}, {Rieke} \& {Luhman}}{{Ivanov} et~al.}{2004}]{Ivanov04}
{Ivanov} V.~D.,  {Rieke} M.~J.,  {Engelbracht} C.~W.,  {Alonso-Herrero} A.,
  {Rieke} G.~H.,    {Luhman} K.~L.,  2004, \apjs, 151, 387

\bibitem[\protect\citeauthoryear{{Kaplan}, {Levine}, {Chakrabarty}, {Morgan},
  {Erb}, {Gaensler}, {Moon} \& {Cameron}}{{Kaplan} et~al.}{2007}]{Kaplan07}
{Kaplan} D.~L.,  {Levine} A.~M.,  {Chakrabarty} D.,  {Morgan} E.~H.,  {Erb}
  D.~K.,  {Gaensler} B.~M.,  {Moon} D.,    {Cameron} P.~B.,  2007, \apj, 661,
  437

\bibitem[\protect\citeauthoryear{{Kenyon}}{{Kenyon}}{1986}]{Kenyon86}
{Kenyon} S.~J.,  1986, {The symbiotic stars}
Cambridge and New York, Cambridge University Press, 1986, 295 p.

\bibitem[\protect\citeauthoryear{{Kleinmann} \& {Hall}}{{Kleinmann} \&
  {Hall}}{1986}]{Kleinmann86}
{Kleinmann} S.~G.,  {Hall} D.~N.~B.,  1986, \apjs, 62, 501

\bibitem[\protect\citeauthoryear{Kramida, {Yu.~Ralchenko}, Reader \& {and NIST
  ASD Team}}{Kramida et~al.}{2013}]{Kramida13}
Kramida A.,  {Yu.~Ralchenko} Reader J.,    {and NIST ASD Team}, 2013, {NIST
  Atomic Spectra Database (ver. 5.1), [Online]. Available:
  {\tt{http://physics.nist.gov/asd}} [2014, January 26]. National Institute of
  Standards and Technology, Gaithersburg, MD.}

\bibitem[\protect\citeauthoryear{{Krivonos}, {Lutovinov}, {Molkov},
  {Revnivtsev}, {Tsygankov} \& {Sunyaev}}{{Krivonos} et~al.}{2013}]{Krivonos13}
{Krivonos} R.,  {Lutovinov} A.,  {Molkov} S.,  {Revnivtsev} M.,  {Tsygankov}
  S.,    {Sunyaev} R.,  2013, The Astronomer's Telegram, 4924, 1

\bibitem[\protect\citeauthoryear{{Kuulkers}, {Shaw}, {Paizis}, {Chenevez},
  {Brandt}, {Courvoisier}, {Domingo}, {Ebisawa}, {Kretschmar}, {Markwardt},
  {Mowlavi}, {Oosterbroek}, {Orr}, {R{\'{\i}}squez}, {Sanchez-Fernandez} \&
  {Wijnands}}{{Kuulkers} et~al.}{2007}]{Kuulkers07}
{Kuulkers} E.,  {Shaw} S.~E.,  {Paizis} A.,  {Chenevez} J.,  {Brandt} S.,
  {Courvoisier} T.~J.-L.,  {Domingo} A.,  {Ebisawa} K.,  {Kretschmar} P.,
  {Markwardt} C.~B.,  {Mowlavi} N.,  {Oosterbroek} T.,  {Orr} A.,
  {R{\'{\i}}squez} D.,  {Sanchez-Fernandez} C.,    {Wijnands} R.,  2007, \aap,
  466, 595

\bibitem[\protect\citeauthoryear{{Luna}, {Sokoloski}, {Mukai} \&
  {Nelson}}{{Luna} et~al.}{2013}]{Luna13}
{Luna} G.~J.~M.,  {Sokoloski} J.~L.,  {Mukai} K.,    {Nelson} T.,  2013, \aap,
  559, A6

\bibitem[\protect\citeauthoryear{{Masetti}, {Dal Fiume}, {Cusumano}, {Amati},
  {Bartolini}, {Del Sordo}, {Frontera}, {Guarnieri}, {Orlandini}, {Palazzi},
  {Parmar}, {Piccioni} \& {Santangelo}}{{Masetti} et~al.}{2002}]{Masetti02}
{Masetti} N.,  {Dal Fiume} D.,  {Cusumano} G.,  {Amati} L.,  {Bartolini} C.,
  {Del Sordo} S.,  {Frontera} F.,  {Guarnieri} A.,  {Orlandini} M.,  {Palazzi}
  E.,  {Parmar} A.~N.,  {Piccioni} A.,    {Santangelo} A.,  2002, \aap, 382,
  104

\bibitem[\protect\citeauthoryear{{Masetti}, {Landi}, {Pretorius}, {Sguera},
  {Bird}, {Perri}, {Charles}, {Kennea}, {Malizia} \& {Ubertini}}{{Masetti}
  et~al.}{2007}]{Masetti07}
{Masetti} N.,  {Landi} R.,  {Pretorius} M.~L.,  {Sguera} V.,  {Bird} A.~J.,
  {Perri} M.,  {Charles} P.~A.,  {Kennea} J.~A.,  {Malizia} A.,    {Ubertini}
  P.,  2007, \aap, 470, 331

\bibitem[\protect\citeauthoryear{{Masetti}, {Munari}, {Henden}, {Page},
  {Osborne} \& {Starrfield}}{{Masetti} et~al.}{2011}]{Masetti11}
{Masetti} N.,  {Munari} U.,  {Henden} A.~A.,  {Page} K.~L.,  {Osborne} J.~P.,
   {Starrfield} S.,  2011, \aap, 534, A89

\bibitem[\protect\citeauthoryear{{Masetti}, {Orlandini}, {Palazzi}, {Amati} \&
  {Frontera}}{{Masetti} et~al.}{2006}]{Masetti06}
{Masetti} N.,  {Orlandini} M.,  {Palazzi} E.,  {Amati} L.,    {Frontera} F.,
  2006, \aap, 453, 295

\bibitem[\protect\citeauthoryear{{McGregor}, {Hart}, {Conroy}, {Pfitzner},
  {Bloxham}, {Jones}, {Downing}, {Dawson}, {Young}, {Jarnyk} \& {van
  Harmelen}}{{McGregor} et~al.}{2002}]{McGregor02}
{McGregor} P.,  {Hart} J.,  {Conroy} P.,  {Pfitzner} L.,  {Bloxham} G.,
  {Jones} D.,  {Downing} M.,  {Dawson} M.,  {Young} P.,  {Jarnyk} M.,    {van
  Harmelen} J.,  2002, SPIE, 4841, 178

\bibitem[\protect\citeauthoryear{{Murset}, {Wolff} \& {Jordan}}{{Murset}
  et~al.}{1997}]{Murset97}
{Murset} U.,  {Wolff} B.,    {Jordan} S.,  1997, \aap, 319, 201

\bibitem[\protect\citeauthoryear{{Nespoli}, {Fabregat} \&
  {Mennickent}}{{Nespoli} et~al.}{2010}]{Nespoli10}
{Nespoli} E.,  {Fabregat} J.,    {Mennickent} R.~E.,  2010, \aap, 516, A94

\bibitem[\protect\citeauthoryear{{Nucita}, {Carpano} \& {Guainazzi}}{{Nucita}
  et~al.}{2007}]{Nucita07}
{Nucita} A.~A.,  {Carpano} S.,    {Guainazzi} M.,  2007, \aap, 474, L1

\bibitem[\protect\citeauthoryear{{Ramirez}, {Depoy}, {Frogel}, {Sellgren} \&
  {Blum}}{{Ramirez} et~al.}{1997}]{Ramirez97}
{Ramirez} S.~V.,  {Depoy} D.~L.,  {Frogel} J.~A.,  {Sellgren} K.,    {Blum}
  R.~D.,  1997, \aj, 113, 1411

\bibitem[\protect\citeauthoryear{{Richichi}, {Fabbroni}, {Ragland} \&
  {Scholz}}{{Richichi} et~al.}{1999}]{Richichi99}
{Richichi} A.,  {Fabbroni} L.,  {Ragland} S.,    {Scholz} M.,  1999, \aap, 344,
  511

\bibitem[\protect\citeauthoryear{{Skrutskie}, {Cutri}, {Stiening} \& {et
  al.}}{{Skrutskie} et~al.}{2006}]{Skrutskie06}
{Skrutskie} M.~F.,  {Cutri} R.~M.,  {Stiening} R.,    {et al.} 2006, \aj, 131,
  1163

\bibitem[\protect\citeauthoryear{{Stacey}, {Heinke}, {Elsner}, {Edmonds},
  {Weisskopf} \& {Grindlay}}{{Stacey} et~al.}{2011}]{Stacey11}
{Stacey} W.~S.,  {Heinke} C.~O.,  {Elsner} R.~F.,  {Edmonds} P.~D.,
  {Weisskopf} M.~C.,    {Grindlay} J.~E.,  2011, \apj, 732, 46

\bibitem[\protect\citeauthoryear{{van Belle}, {Lane}, {Thompson}, {Boden},
  {Colavita}, {Dumont}, {Mobley}, {Palmer}, {Shao}, {Vasisht}, {Wallace},
  {Creech-Eakman}, {Koresko}, {Kulkarni}, {Pan} \& {Gubler}}{{van Belle}
  et~al.}{1999}]{vanBelle99}
{van Belle} G.~T.,  {Lane} B.~F.,  {Thompson} R.~R.,  {Boden} A.~F.,
  {Colavita} M.~M.,  {Dumont} P.~J.,  {Mobley} D.~W.,  {Palmer} D.,  {Shao} M.,
   {Vasisht} G.~X.,  {Wallace} J.~K.,  {Creech-Eakman} M.~J.,  {Koresko} C.~D.,
   {Kulkarni} S.~R.,  {Pan} X.~P.,    {Gubler} J.,  1999, \aj, 117, 521

\bibitem[\protect\citeauthoryear{{van den Berg}, {Grindlay}, {Laycock}, {Hong},
  {Zhao}, {Koenig}, {Schlegel}, {Cohn}, {Lugger}, {Rich}, {Dupree}, {Smith} \&
  {Strader}}{{van den Berg} et~al.}{2006}]{vandenBerg06}
{van den Berg} M.,  {Grindlay} J.,  {Laycock} S.,  {Hong} J.,  {Zhao} P.,
  {Koenig} X.,  {Schlegel} E.~M.,  {Cohn} H.,  {Lugger} P.,  {Rich} R.~M.,
  {Dupree} A.~K.,  {Smith} G.~H.,    {Strader} J.,  2006, \apjl, 647, L135

\bibitem[\protect\citeauthoryear{{van den Berg}, {Penner}, {Hong}, {Grindlay},
  {Zhao}, {Laycock} \& {Servillat}}{{van den Berg} et~al.}{2012}]{vandenBerg12}
{van den Berg} M.,  {Penner} K.,  {Hong} J.,  {Grindlay} J.~E.,  {Zhao} P.,
  {Laycock} S.,    {Servillat} M.,  2012, \apj, 748, 31

\bibitem[\protect\citeauthoryear{{Wallace} \& {Hinkle}}{{Wallace} \&
  {Hinkle}}{1997}]{Wallace97}
{Wallace} L.,  {Hinkle} K.,  1997, \apjs, 111, 445

\end{thebibliography}

\bsp

\label{lastpage}

\end{document}